\documentclass[sigconf,nonacm]{acmart}
\usepackage{tikz}
\usepackage{listings}
\usepackage{todonotes}
\usepackage{subcaption}
\usepackage{enumitem}
\usepackage{pgfplots}
\pgfplotsset{compat=1.3}

\usepackage[norelsize, linesnumbered, ruled, lined, boxed, commentsnumbered]{algorithm2e}

\newcommand\Tmall{\mathcal{T}_{M}}
\newcommand\Tmc{\mathcal{T}_{m}}
\newcommand\aTmc{\widetilde{\mathcal{T}}_{m}}
\newcommand\Tsall{\mathcal{T}_{S}}
\newcommand\Tsim{\mathcal{T}_{s}}
\newcommand\aTsim{\widetilde{\mathcal{T}}_{s}}
\newcommand\aT{\widetilde{\mathcal{T}}}

\newtheorem{problem}{Problem}

\AtBeginDocument{%
  }

\setcopyright{acmlicensed}
\copyrightyear{2018}
\acmYear{2018}
\acmDOI{XXXXXXX.XXXXXXX}

\acmConference[Conference acronym 'XX]{Make sure to enter the correct
  conference title from your rights confirmation emai}{June 03--05,
  2018}{Woodstock, NY}
\acmISBN{978-1-4503-XXXX-X/18/06}

\begin{document}


\title{Formal Model Guided Conformance Testing for Blockchains}

\author{Filip Drobnjakovi\'c}
\author{Matija Kupre\v sanin}
\author{Pavle Suboti\'c}
\email{filip,matija,psubotic@fantom.foundation}
\affiliation{%
  \institution{Sonic Research}
  \country{Serbia}
}

\author{Amir Kashapov}
\author{Bernhard Scholz}
\email{amir,bernhard@fantom.foundation} 
\affiliation{%
  \institution{Sonic Research}
  \country{Australia}
}

 \titlenote{all authors contributed equally to this research.}



\begin{abstract}

 Modern blockchains increasingly consist of multiple clients that implement a single blockchain protocol. If there is a semantic mismatch between the protocol implementations, the blockchain can permanently split and introduce new attack vectors.  Current ad-hoc test suites for client implementations are not sufficient to ensure a high degree of protocol conformance. As an alternative, we present a framework that performs protocol conformance testing using a formal model of the protocol and an implementation running inside a deterministic blockchain simulator. Our framework consists of two complementary workflows that use the components as trace generators and checkers. Our insight is that both workflows are needed to detect all types of violations. We have applied and demonstrated the utility of our framework on an industrial strength consensus protocol. 


\end{abstract}

\keywords{Testing, Conformance, Blockchain, Formal Verification}

\received{20 February 2007}
\received[revised]{12 March 2009}
\received[accepted]{5 June 2009}

\maketitle

\section{Introduction}
\label{sec:introduction}
Blockchains are increasingly being utilized across various industries. Blockchains offer a decentralized computing platform that increases trust, security and transparency by maintaining a public ledger of all transactions. This is achieved by several client nodes communicating with each other using a uniform protocol. While recent market analyses~\cite{deloite} have highlighted blockchain’s transition from an emerging technology to an integral part of the global digital infrastructure, the robustness~\cite{hacks, hacks3, netpart} of blockchains remains a fundamental concern that needs to be addressed for continued adoption. 


To improve robustness, notable blockchains such as Ethereum have resorted to the use of several heterogeneous client~\cite{ethclients} implementations. These client implementations are often written in different programming languages~\cite{nversionp} to increase client diversity. For example, Ethereum has 6 client implementations written in 5 different programming languages~\cite{ethclients}. The hope is that with a diverse set of clients, a fault in a particular implementation is unlikely to manifest in other clients~\cite{ethclients}.

While multi-version client blockchains indeed improve robustness from implementation bugs, they present another challenge. Having several blockchain client implementations can be disastrous if their implementations do not follow a single protocol e.g., consensus protocols. This problem was evidenced in 2024 when different versions of the Ethereum client~\cite{gethblog} caused a debilitating network partition. Such incidents occur because the correct functionality of blockchains inherently assumes all nodes are operating with the same protocol. While in-built byzantine fault tolerance~\cite{BFT} allows the protocol to tolerate specific byzantine behaviors (failure and equivocation), blockchains with incompatible protocols can permanently split and  introduce new vulnerabilities~\cite{doublespending, attackseth}. 

Currently, Ethereum provides test suites~\cite{ethtests} to provide lightweight conformance testing. However, ad hoc single client test cases are not adequate for protocol conformance testing. Firstly, in documents, descriptions of protocols are often ambiguously written and lack rigor or formalization, which makes it difficult to determine the set of test cases that describe a specific protocol. Moreover, unlike sequential non-distributed systems, the properties blockchain protocols must adhere to are \emph{network-centric}\footnote{also called global properties or system-wide properties.} as opposed to \emph{node-centric}. Therefore, testing single nodes independently using test suites cannot capture important properties e.g. consensus consistency, which require the collective state of all clients. Lastly, testing nodes on a live network is impractical. Without determinism and a mechanism to inject faults, it may take months or years for incorrect behavior to occur that is reproducible.


\begin{figure}
    \centering
    \includegraphics[width=0.99\linewidth]{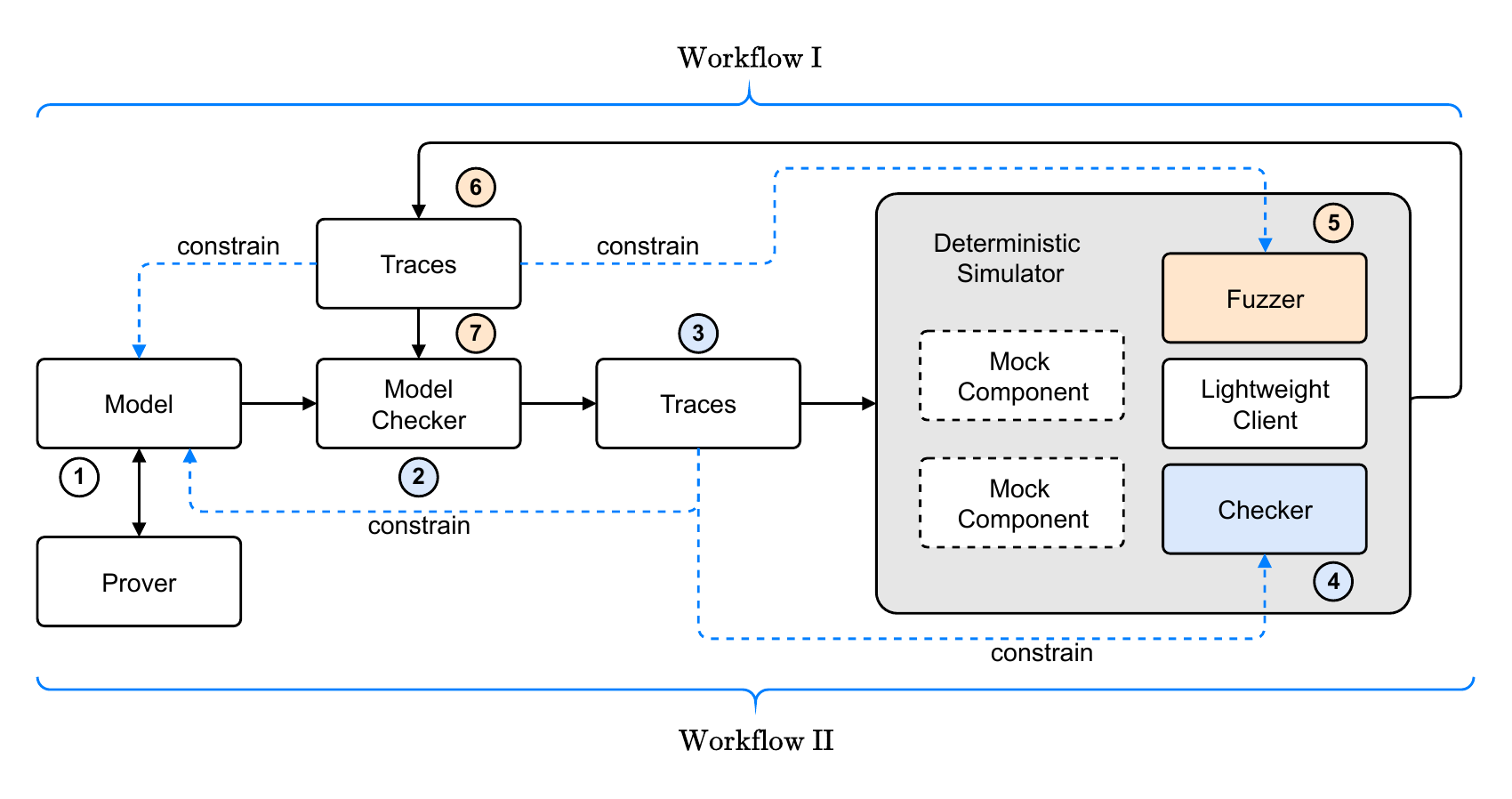}
    \caption{Formal Model Deterministic Simulation Environment}
    \label{fig:FMDD}
\end{figure}

To address the above challenges, we propose a new testing framework called \emph{Formal Model Deterministic Simulation Environment} (FMDSE). Our framework centers around a formal system model and a deterministic blockchain simulator.  The formal model establishes the conceptual design of a blockchain component, e.g., consensus protocol, using a formal logical language and we \emph{prove} safety properties hold for all scenarios in the model. To establish that the implementation has a high degree of conformance to the model, we generate and check traces from both the model and implementation. Using a model checker, we generate random traces from the model and check the implementation results are equivalent in the final and intermediate states of each trace. Symmetrically, we randomly generate traces in the implementation and use a model checker to ensure they 
are contained in the set of traces described by the model. However, due to the fact that distributed systems such as blockchains are non-deterministic, we do not deploy the blockchain implementation in the real world directly. Instead, we use a deterministic blockchain simulator to control the execution of the implementation in a simulated environment that mocks and assigns fixed values to random/unpredictable aspects such as a network latency, clock, randomness beacons, equivocation etc. The deterministic simulator also allows
us to obtain the network-centric state of the blockchain directly (with no network communication) and by the use of virtual time, speed up testing. We describe our framework in Figure~\ref{fig:FMDD}. 

Our framework consists of two complementary workflows that can be combined as a feedback loop, used sequentially or a variation of the two. For both workflows, we require a formal model \textcircled{1} which specifies network-wide behaviors and properties. For instance, in TLA+~\cite{Lamport-book2002, tla} this would be represented as a state transition system using the mathematical logic-based TLA+ language. We prove the specification safe\footnote{by finding an inductive invariant.}  with respect to its properties using a theorem prover~\cite{CDLMRV-fm12} or automated reasoner~\cite{swiss}. Since we define the model as a state transition system, we can view its behavior as a set of traces. However, since the formal model is defined at a higher level of abstraction than the implementation, we require a mechanism to establish if an execution in the implementation is correct with respect to the model. To do so, in Workflow I (\textcircled{5}, \textcircled{6}, \textcircled{7}), we generate random traces \textcircled{6} from the simulator by fuzzing input parameters \textcircled{5} and use the model to check that these traces correspond to the model \textcircled{7} using a model checker. Each time a model trace passes (is correct) in the simulator, we constrain the model and simulator to not generate this trace in the future. In Workflow II (\textcircled{2}, \textcircled{3}, \textcircled{4}), we generate random model traces \textcircled{3} from a model checker \textcircled{2} and check that the simulator produces the same abstract states as the model \textcircled{4}. Again, each time a simulator trace passes in the model, we constrain both the model and simulator to not generate this trace in the future. It is important to note that both workflows are required and complementary. In Workflow I, if a trace from the simulator cannot be validated by the model, then either the implemented protocol is over-approximating and allows unintended behavior (possibly erroneous) or the model has not foreseen such an execution and needs refinement. In Workflow II, if a model trace cannot be executed by the simulator then the implementation of the protocol is too restrictive or the model is too unconstrained. 
In the case that any of the traces are correct, 
we constrain the complementary source to not generate the same traces. The main insight of our technique is that both workflows are necessary to detect all types of conformance violations. That is, Workflow I can detect violations Workflow II cannot, and vice versa.

We have implemented the FMDSE framework and used it to check conformance of an asynchronous consensus protocol implementations for the \textsc{Sonic} blockchain. A consensus protocol is a key component in the blockchain client software that ensures that all nodes in a blockchain produce the same block of transactions. These protocols are very subtle, in that small changes to the design can quickly result in a loss of the safety i.e., consistency, rendering any potential performance improvements obsolete. Aside from evaluating the conformance of consensus protocol implementations, the FMDSE has aided us in understanding the intricate details of protocols that have been a catalyst for novel optimizations. We plan on replicating this development approach to other components of the \textsc{Sonic} blockchain in the future. 


We summarize our contributions as follows.
\begin{itemize}
    \item A novel conformance testing framework for blockchains. Our approach combines deterministic simulation with a formal model, to test a given implementation conforms to a formal model. Compared to the state-of-the-art, our approach is bi-directional, performing violation detection in two complementary directions resulting in improved coverage. On the other hand, by utilizing fuzzing, we are able to scale to large-sized specifications. 
    \item An implementation of our FMDSE framework that has been used to conformance test client nodes in the \texttt{Sonic} blockchain.
    \item An industrial evaluation including a detailed reflection of lessons learnt while developing and testing with the FMDSE. 
\end{itemize}

This paper is structured as follows. In Section~\ref{sec:usecase} we describe our use case that motivates our technique.  In Section~\ref{sec:bugs} we motivate our two workflows by defining the violations they can detect. In Section~\ref{sec:formal} we describe how we specify and prove our formal model in Section~\ref{sec:sim} we describe our deterministic simulation environment, and we present the general algorithm with variations in Section~\ref{sec:testing}. In Section~\ref{sec:evaluation} we provide an experimental evaluation, in Section~\ref{sec:dis} we provide a discussion of our work and present related work in Section~\ref{sec:relatedwork}. We conclude in Section~\ref{sec:conclusion}.

\section{Use Case Description}
\label{sec:usecase}

In this section, we describe our use case, namely, the development of blockchain consensus protocols.

\begin{figure}
    \centering
    \includegraphics[width=0.4\linewidth]{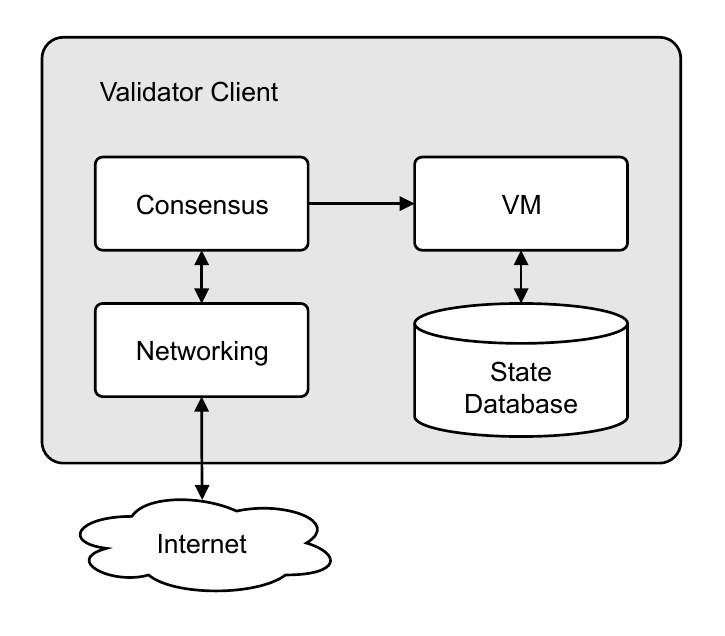}
    \caption{Node Client Software}
    \label{fig:client}
\end{figure}

A blockchain is a network of nodes that keeps
a shared ledger. The ledger is constructed via a chain of blocks where each block contains transactions and is cryptographically secured. Each node runs client software that has the architecture depicted in Figure~\ref{fig:client}. Transactions are disseminated to other nodes over the Internet governed by a network component. These transactions are ordered and packed into a \emph{block} by the consensus component. Each block is executed by a \emph{virtual machine}, while maintaining a data store i.e., state database component for managing account balances on the ledger.

A vital property of blockchains is that each of the nodes agree with each other over what the next block is. Failure to do so can result in security vulnerabilities such as double spending~\cite{doublespending}. 

This agreement is achieved by a \emph{consensus protocol}, implemented in the consensus component of the client software. The consensus protocol is an algorithm that determines how each node should share information so that 
they reach an agreement on the next block. The consensus protocol is considered \emph{safe} if it is always the case that all nodes agree with each other on the next block. We say it is \emph{live}, if all participants are guaranteed to reach consensus.  To add robustness, consensus protocols typically assume up to 1/3 byzantine faults (exclusive)~\cite{BFT} i.e., nodes that can fail and/or send different duplicating transactions to other nodes. 

\begin{example}[DAG Consensus Protocol]
We present a consensus protocol where each node stores a \emph{layered}
Directed Acyclic Graph (DAG). We show a snapshot of a consensus protocol DAG in Figure~\ref{fig:consensus}.  In the network there are $3$ nodes, labeled $P1$, $P2$, and $P3$. Each local DAG depicts the local view of the node, as shown in Fig.~\ref{fig:dag}. In the DAG, each vertex represents a transaction\footnote{or a list of transactions} and an edge represents a happens-before relationship~\cite{lamport-time}.  The DAG is layered by rounds (rows), starting at round 0  to some integer $N$ so that each vertex in a given row is in the same round. The vertices in previous rounds, that are linked to by a vertex are called parents of that vertex. 

The protocol roughly works as follows: each node is assigned a stake (an integer value). Each node creates its own initial vertex in round $0$ and broadcasts it to the other nodes. It listens for incoming vertices and when it receives 2/3 of the total stake in a round, it creates a new vertex in the next round and links to the vertices (parents) of the previous round. At every odd round, a leader vertex is chosen by some global criteria. At every even round, the vertex is finalized if $>$2/3 of the stake can reach it. All vertices between two leaders can be deterministically linearized via topological sort to form a block of transactions as depicted in Figure~\ref{fig:lin}.

As the blockchain runs, the DAG grows in an upwards manner. Some local views of a DAG may grow faster than others but the DAG will be consistent: every vertex with the same round and creator node, irrespective of what local DAG it is in, will always have the same parents. Given that, it follows that there will be eventual consistency between local DAGs.


\begin{figure}
	\centering
	\begin{subfigure}{0.9\linewidth}
	    \includegraphics[width=0.99\linewidth]{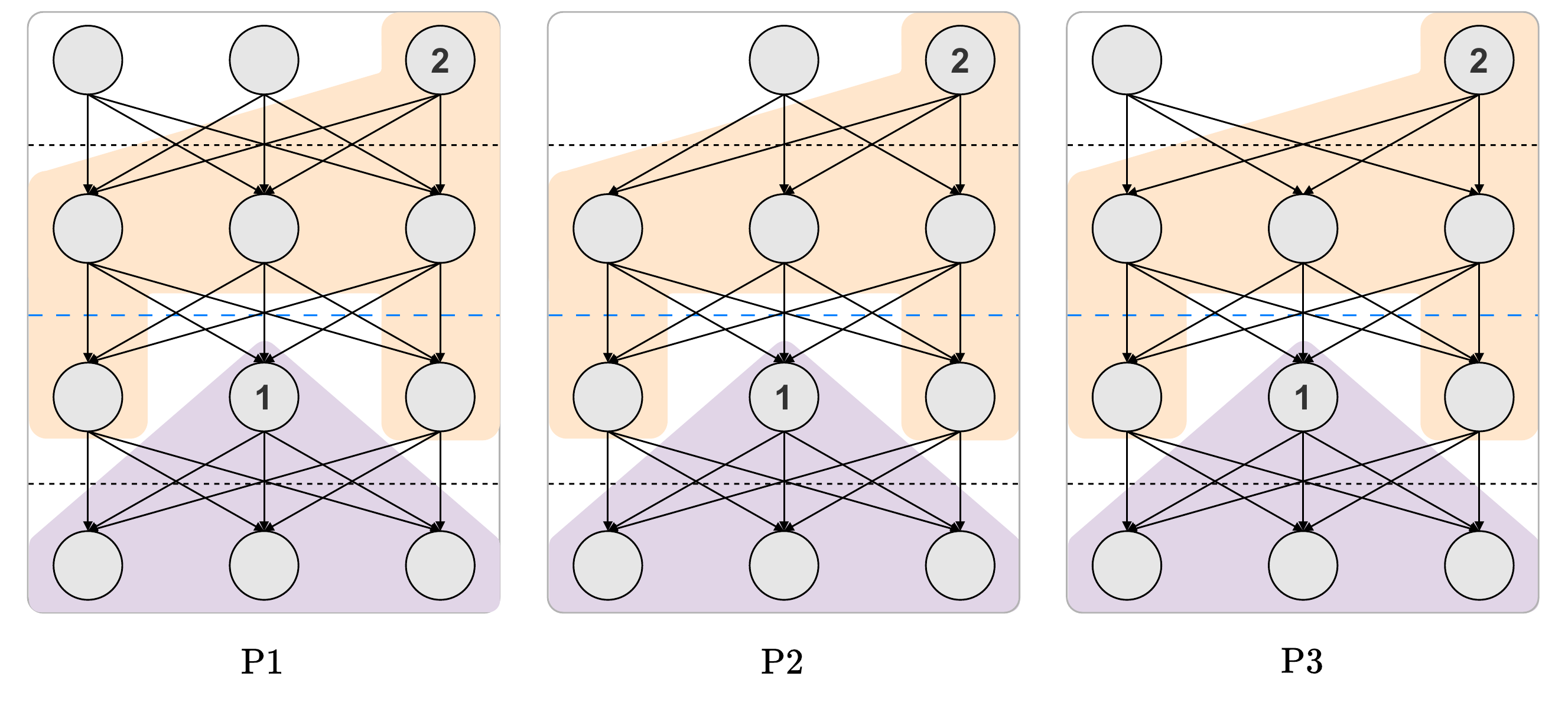}
		\caption{Snapshot of local DAGs of nodes P1, P2 and P3}
		\label{fig:dag}
	\end{subfigure}\hfill
	\begin{subfigure}{0.7\linewidth}
		\includegraphics[width=0.99\linewidth]{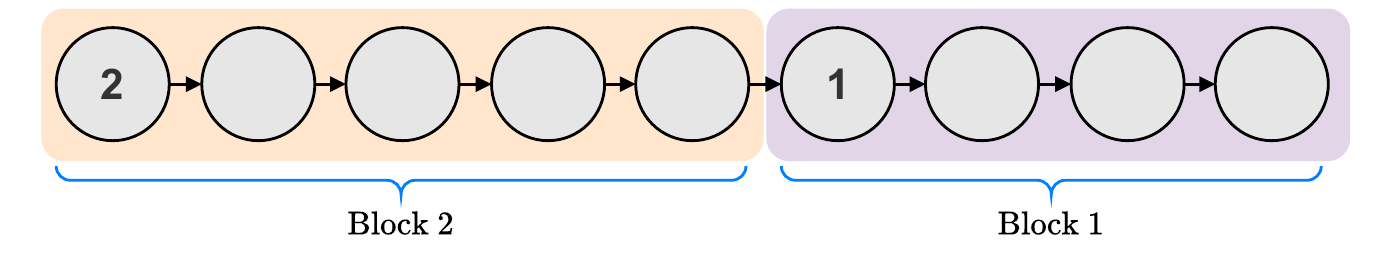}
		\caption{Blocks formed from linearizing DAG}
		\label{fig:lin}
	\end{subfigure}
	\caption{Simplified DAG Consensus Protocol}
	\label{fig:consensus}
\end{figure}

\end{example}

When we describe a consensus protocol we assume that all nodes in the blockchain operate with the same protocol. Even slight deviations can cause network splits or inconsistency. However, the line between optimizations and 
divergence from the protocol can be blurred when the specification is ambiguously written in a document or worse, taught by word of mouth from engineer to engineer. Therefore, the protocol conformance problem is vital when several clients operate in a single blockchain. 

\begin{problem}[Protocol Conformance in a Blockchain]
    Given $N$ client implementations and given a protocol specification $S$, 
    do all $N$ nodes conform to the specification $S$?  
\end{problem}

\begin{example}[Incompatible DAG Consensus Protocols]
Consider two DAG protocol implementations where the client picks a leader based on processor ID.  While developing one of the implementations, a new developer might observe that this can lead to bias against high-stake nodes and decide to make a new client that selects the leader based on stake. If for example, the current stake in the blockchain is ordered $P3 > P2 > P1$. This will cause inconsistency in the choice of anchor vertex selected across nodes and as a consequence inconsistency with the blocks produced by each node, opening up attack vectors~\cite{doublespending} in the blockchain.
\end{example}



\section{Conformance Violations}
\label{sec:bugs}
In this section, we describe the types of conformance violations we aim to detect with our framework. We categorize violations into 3 categories: Type-I, Type-II and Prop. A single violation can belong to multiple categories, depending on the workflow used. However, a violation can also belong to only one of these categories, thus our categorization is important for arguing why both Workflow I and II are required to detect all conformance violations. Prop violations cannot be classified as any other type. Assuming a violation is not that of a property: if a violation can be detected by Workflow I it can be classified as Type-I; similarly, if it can be detected by Workflow II it can be considered Type-II.

\paragraph{\bf Preliminaries}
Recall that our framework uses traces generated from a formal model and a simulated implementation to perform conformance testing. The model and simulated implementation can be described as a \emph{labeled transition system} (LTS).  A labeled transition system is a tuple $(S, \Lambda, Tr)$ where $S$ is a set of states,  $\Lambda$ is a set of labels (which we refer to as actions), and $Tr$ the labeled transition relation is a subset of $S\times \Lambda \times S$. We say that there is a transition from state $p$ to state $q$ with action $a$ iff $(p, a, q) \in Tr$ and denote it $p \xrightarrow{a} q$. Let the simulated implementation be represented by the transition system: $(S_{s}, \Lambda_{s}, Tr_{s})$ and the model be represented by a transition system: $(S_{m}, \Lambda_{m}, Tr_{m})$. An execution of a transition system is represented by a \emph{trace}. A trace is a non-empty sequence of program states and actions that respects the transition relation $Tr$, that is, $(s, a, s') \in Tr$ for each pair of consecutive states $s, s' \in S$ in the sequence. Thus, we can describe a trace with the following notation:
$s_0 \xrightarrow{a_i} s_1 \xrightarrow{a_j} s_2 \xrightarrow{a_k} \dots $ where $a_i, a_j, a_k \in \Lambda$. 

We let $\Tmc$ be the set of (correct) traces that can be generated from the model and $\Tmall$ the set of all correct and incorrect traces such that $\Tmc \subseteq \Tmall$. Likewise, let $\Tsim$ be the set of (correct) traces that can be generated from the simulated implementation and $\Tsall$ the set of correct and incorrect traces such that $\Tsim \subseteq \Tsall$.  In the formal model we define a property $\Phi$ such that $\Tmc \models \Phi$ i.e., $\forall t \in \Tmc: t \models \Phi$.

Since the model defines the protocol at a much more abstract level than the implementation i.e., the implementation may have more variables, several actions for one model action, etc. we define a mapping function $\Xi$ that maps implementation traces to model traces so that $\Xi(t) \in \Tmall$ where $t \in \Tsim$. We also define $\theta$ that maps model traces to a set of simulator traces such that $|\theta(t)| \ge 1$ where $t \in \Tmc$ and $\theta(t) \subseteq \Tsall$. $\Xi(t)$ takes a trace $s_0 \xrightarrow{a_i} s_1 \xrightarrow{a_j} s_2 \xrightarrow{a_k} \dots s_n$ and produces a trace that 
reduces each state $s$ with variables only in the more abstract 
model LTS, concatenates several actions that can be represented by a single action in $\Lambda_m$ and removes unneeded intermediate states. $\theta(t)$ does the inverse, given a 
model trace $s_0 \xrightarrow{a_i} s_1 \xrightarrow{a_j} s_2 \xrightarrow{a_k} \dots s_n$ it generates several traces by replacing each trace step $(s, a, s')$ with several subtraces that can lead from an equivalence class of states $s$ to $s'$. Given these definitions, we define an abstraction and concretization relations on sets of traces as follows.


\begin{align}
    \alpha(T) &= \{ \Xi(t) \ | \ t \in T\}\\
   \gamma(T) &= \bigcup_{t \in T} \theta(t)
\end{align}

Here $\alpha$ takes the set of traces that the simulated implementation produces and converts them into model traces. $\gamma$ takes the set of 
traces produced by the model and produces a set of traces that the simulated implementation can ingest. 

However, since we perform fuzzing, we define $\aTmc$ to be the corresponding finite set resulting from the approximation of $\Tmc$ by model checker fuzzing (Workflow II) such that $\aTmc \subseteq \Tmc$ (in practice $\aTmc \subset \Tmc$). Similarly, given $\Tsim$ is the allowed behavior of the simulated implementation,  we define $\aTsim$ to be the set of traces of the that approximate $\Tsim$ by fuzzing (Workflow I) such that $\aTsim \subseteq \Tsim$ (in practice $\aTsim \subset \Tsim$). Note that it does not necessarily hold that $\alpha(\Tsim) \models \Phi$ even if $\alpha(\aTsim) \models \Phi$.  

With the above definitions in mind, we categorize the type of violations that we aim to detect as follows. 

\paragraph{\bf Protocol Conformance Type-I}
We identify a conformance violation Type-I that occurs when the approximated behavior in the simulator does not break a property i.e., $\alpha(\aTsim) \models \Phi$, but a corresponding behavior does not exist in the model i.e., $\exists T \subseteq \alpha(\aTsim) : T \neq \emptyset \wedge (T \cap \Tmc = \emptyset)$. This indicates that the implementation does not fully correspond to the model and either needs to be constrained or the model needs to be unconstrained.  

\begin{example}[Vertex Creation Semantics]
In the consensus protocol in Figure~\ref{fig:dag}, a node can create a vertex only after receiving a quorum of vertices from another node. However, this is a very subtle protocol detail that can be missing during implementation. If the implementation allows unconstrained vertex creation, it can starve other nodes and lead to liveness issues. This violations cannot be detected by Workflow II. 
\end{example}


\paragraph{\bf Protocol Conformance Type-II}
We identify a conformance violation Type-II that occurs when the approximated behavior of the model doesn't break any properties i.e., $\aTmc \models \Phi$, but cannot be simulated i.e., $\exists T \subseteq \aTmc : T \neq \emptyset \wedge (T \cap \alpha(\Tsim) = \emptyset)$. This is the dual of Type-I violations; indicating the implementation is too constrained or the model is too unconstrained. 

\begin{example}[Byzantine Behavior]
In consensus protocols such as the one described in Figure~\ref{fig:dag},  there can be nodes that are called Byzantine. These nodes can create two vertices at once and 
try to trick the other nodes in the blockchain to have inconsistent DAGs. An implementation that doesn't account for such behavior can be still safe and tolerate less than half Byzantine nodes. However, given a trace that does equivocate it may end up in an inconstant state that will diverge from the state of the trace from the model. This type of violation can be detected by Workflow II but not Workflow I.
\end{example}


\paragraph{\bf Violation of Property}
The last violation type is one where $\alpha(\aTsim) \not\models \Phi$. A trace in $\alpha(\aTsim)$ can be detected by both workflows, directly, by the simulator producing a trace $t \in \aTsim$ where $\alpha(t) \not\models \Phi$ or indirectly by the model checker producing a correct trace $t \in \aTmc$ such that $\aTmc \models \Phi$ but leads the simulator to produce a counter-example trace $cex$ such that $\alpha(\{cex\}) \in \Tmall \setminus \Tmc$ because $\alpha(\{cex\}) \not \models \Phi$. 

\begin{example}[Data Structure Assumptions]
Consider that the DAG in Figure~\ref{fig:dag} needs to be linearized once a leader is found, as indicated by the shaded regions. An implementation may assume a data structure such as a map in golang that orders vertices by node id for each round. However, if that map data structure is unordered, it will lead to inconsistent linear orders and result in an unsafe blockchain. 
\end{example}

\section{Formal Models}
\label{sec:formal}
In this section we describe the development of formal models in our framework. 

The development of formal models can be seen in the software engineering lifecycle as a design step to rigorously specify and prove the intended behavior of critical components. In addition to the rigorous proving, formal models allow the conceptual design of the 
component to be iterated on without being bogged down with implementation details that are abstracted in the model. 

For our use case, namely, the consensus protocol, this is an important step for a number of reasons. Consensus protocols are notoriously difficult to get right. An incorrect consensus protocol at the conceptual level will cause failure no matter how well optimized it is.  At the same time, the specification should not constrain the design too much or it may inhibit any performance gains in the implementation. 

Similar to a design document, the specification also acts as a source of truth on how the protocol should operate and what properties it must adhere to. With multiple client implementations becoming ever more popular in blockchains, a single authoritative source of truth is important for the various implementations to correctly co-exist.  

In our use case, to specify our consensus protocol, we use TLA+ as a specification language. TLA+\cite{Lamport-book2002} is a language / toolbox to specify systems, notably concurrent and distributed algorithms. It is based on Zermelo-Fraenkel set theory with choice for representing the data structures on which the algorithm operates, and the Temporal Logic of Actions (TLA), a variant of linear-time temporal logic (that allows binary \emph{action} relations on states), for describing executions of the algorithm.

\paragraph{\bf Specifications} To specify a system in TLA+ we use \emph{state variables} and \emph{actions}. A system state is an assignment of values to state variables.  
An action is a binary relation on states, specifying the effect of executing a sequence of instructions. 
An action is represented by a formula over unprimed and primed variables where unprimed variables refer to the values of the variables in the current state and primed variables refer to the values of the variables in the  resulting state. For example, the instruction $x := x+1$ is represented in TLA+ by the action $\texttt{x' = x+1}$.  A system is specified by its actions and initial states, i.e., $\texttt{Spec} = \texttt{Init} \land \Box\texttt{[Next]}_{\texttt{vars}}$ where \texttt{Init} is the initial states predicate, and \texttt{Next} is a disjunction of all actions of the system, and \texttt{vars} is the tuple of all state variables. The expression $\texttt{[Next]}_{\texttt{vars}}$ is true if either \texttt{Next} is true, meaning that some action is true and therefore executed, or \texttt{vars} stutters, meaning that the values of the variables are the same in the current and next states. The symbol $\Box $ is the temporal operator ``always''. Thus, \texttt{Spec} defines a set of infinite sequences of system states, i.e., those such that (1) the first state satisfies \texttt{Init}, and (2) every successive pair of states satisfies $\texttt{[Next]}_{\texttt{vars}}$. Such a sequence is called a \emph{behavior}.

\paragraph{\bf Properties} Given a TLA+ specification, one can then define safety properties of the set of behaviors. A safety property  in TLA+ is a formula of the form $\texttt{Spec => } \Box \texttt{Inv}$, meaning that every state on every behavior that satisfies the system specification, satisfies the state predicate $\texttt{Inv}$, a formula over the (unprimed) state variables. We call the predicate $\texttt{Inv}$ a \emph{safety invariant}. 

\paragraph{\bf Proofs} To prove safety, one needs to find a safety invariant that is \emph{inductive}. An inductive invariant $Ind$ is an invariant that is typically
stronger than the desired safety invariant and is preserved
by all protocol transitions. Thus, $Ind$ satisfies the following conditions:
\begin{align}
     &Init \implies Ind \\
     &Ind \wedge Next \implies Ind' \\
     &Ind \implies Inv 
\end{align}

Here, where $Ind'$ denotes the predicate $Ind$ 
where state variables are replaced by their primed next-state versions. Conditions (1) and (2) are, respectively, referred to as initiation and consecution. Condition (1) states that $Ind$ holds at all initial states. Condition (2) states that $Ind$ is inductive, i.e., if it holds at some state $s$ then it also holds at any successor of $s$. Together these two conditions imply that $Ind$ 
is also an invariant, i.e., that it holds at all reachable states. Condition (3) states that $Ind$ is possibly stronger than the invariant $Inv$ that we are trying to prove. Therefore, if all reachable states satisfy $Ind$, they also satisfy $Inv$. The challenge is to find $Ind$ which is typically performed with a semi-automated theorem prover~\cite{tlaps} or an automated reasoner~\cite{swiss, SchultzDT22}. We note that the specification must have some mapping to the implementation for our technique to work. Generally the specification is more abstract; State variables in the specification will exist in the implementation but they will select efficient data structures etc. However, many variables in the implementation will not exist in the specification. In the implementation several actions may map to one action in the specification; nevertheless, this mapping must exist.



\lstset{language=Python, basicstyle=\scriptsize\ttfamily, breaklines=true, backgroundcolor=\color{gray!10}, showstringspaces=false,
frame=ltb,
framerule=0pt,
}

\begin{example}[Simple DAG Protocol Specification in TLA+] Consider the simplified DAG protocol in Section~\ref{sec:usecase}. The core of the state is the DAG that holds the causal information
of the vertices. We describe the DAG in Listing~\ref{lst:dag}. Note, the definition is network-centric. It is first indexed by \texttt{NodeSet} i.e., the set of nodes in the 
blockchain to specify which local view of the DAG we want to access. Next, for that local view we index by \texttt{RoundSet} i.e., the set of rounds in the DAG, to indicate which round we are interested in. Finally, 
we index again by \texttt{NodeSet} to indicate the creator of the vertex. 

\begin{lstlisting}[label=lst:dag, caption=Network centric DAG state] 
dag \in [
        NodeSet->[
                    RoundSet->[
                                NodeSet->VertexSet \cup NilVertexSet
                    ]
        ]
    ]
\end{lstlisting}

The \texttt{Init} and \texttt{Next} transitions are defined below in Listing~\ref{lst:transitions}. Then \texttt{Init} sets the initial state for the 
data structures. For example \texttt{InitDag} 
initializes the dag state variable to an a mapping 
to \texttt{NilVertexSet}. The \texttt{Next} assumes an initialized state and specifies that 
the state transition system for a node \texttt{p} in \texttt{NodeSet} has a transition  
\texttt{ReceiveVertexTn(p, q, r, v)} to receive a vertex for some 
round from another node \texttt{q} in \texttt{NodeSet}. The transition system for node 
\texttt{p} can also transition to the next round 
if the condition \texttt{NextRoundTn(p)} is satisfied.

\begin{lstlisting}[label=lst:transitions, caption=Protocol Transitions]
Init == 
   /\ InitDag
   /\ InitLeader
   ...

Next == 
   \E p \in NodeSet, r \in RoundSet, v \in VertexSet: 
      \E q \in NodeSet\{p}:
         \/ NextRoundTn(p)
         \/ ReceiveVertexTn(p, q, r, v)
         ...
\end{lstlisting}

In Listing~\ref{lst:prop} we specify the safety property, which states that under our specification, leaders remain constant across nodes and are monotonically increasing at all nodes.

\begin{lstlisting}[label=lst:prop, caption=Safety Property]
Safety == Spec => [](LeaderConsistency /\ LeaderMonotonicity)
\end{lstlisting}

We can prove the specification semi-automatically using the TLAPS~\cite{tlaps} proof assistant for TLA+ or with an invariant discovery tool~\cite{SchultzDT22}.
\end{example}

\section{Deterministic Simulation}
\label{sec:sim}
In this section, we describe our deterministic blockchain simulator. The goal of the deterministic blockchain simulator is to control uncertain parameters present in a real world blockchain and to allow us to capture the network-wide state from multiple node instances while enabling rapid, reproducible testing.

Our simulator is designed so that the software component under test, the consensus protocol, remains agnostic to operation in a real client or simulated environment. This is achieved through the abstraction of fundamental components that introduce non-determinism in distributed systems such as the clock and network communication. These abstractions are exposed through interfaces that allow for multiple implementations, each serving different testing objectives while maintaining deterministic behavior.

The simulator implements a Discrete Event Simulator (DES) that models system operation as a sequence of events in time, where each event marks a state transition at a specific instant. Between consecutive events, no state changes occur, allowing simulation time to advance directly to the next event through next-event time progression. This model enables our reference clock implementations: a simulation clock that accelerates execution by advancing directly between state transitions, and a system clock that aligns with real-time progression. Additional clock models, such as those with specialized timing behaviors, can be integrated through the same interface while maintaining deterministic execution. The DES architecture allows for precise logging and analysis of state transitions, enabling targeted identification of bugs through exact traces rather than relying on chance encounters during extended test runs. Through this approach, test scenarios that would require hours or days of system time can be executed within seconds or minutes of simulation time, substantially reducing the testing cycle while maintaining precision and reproducibility.

Our network abstraction supports two reference implementations designed for distinct testing scenarios: a simulated network that enables control over message delivery delays, peer connectivity, network faults, and other parameters, and a fake network that provides instant, lossless message delivery between fully connected nodes for isolating consensus behavior from network effects. These implementations demonstrate how additional network models, such as those for Byzantine behaviors or specific network topologies, can be integrated through the same interface.

Lastly, to maintain this deterministic behavior, our simulator eliminates sources of randomness and non-determinism. For instance, we avoid multithreading, and where protocols may require randomness, we employ pseudorandom number generators with reproducible results via consistent seeds. By removing these variables along with physical layer dependencies, we ensure complete reproducibility of all test scenarios and protocol behaviors. The combination of configurable network conditions and deterministic execution through discrete transitions guarantees that identified bugs can be reproduced, diagnosed, and fixed.

\begin{figure}
    \centering
    \includegraphics[width=0.77\linewidth]{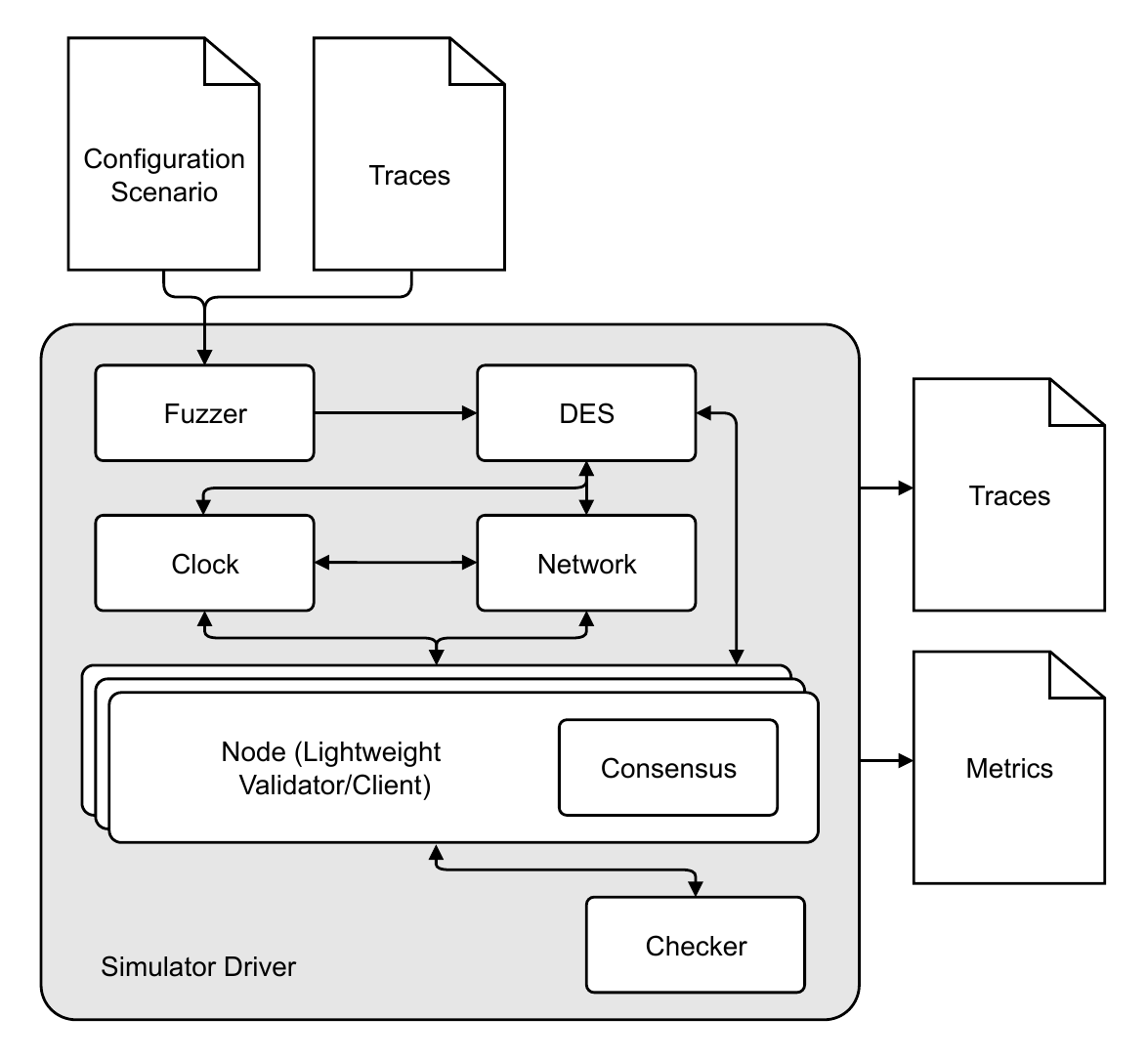}
    \caption{Deterministic Blockchain Simulator \label{fig:sim}}
\end{figure}

Figure~\ref{fig:sim} illustrates the structure of the deterministic simulator to test a consensus protocol. The simulator allows for instantiation of one or more nodes (lightweight node abstractions) connected to other external components such as the mocked network and virtual clock. In the case of Workflow I, the user provides ranges of parameters via the configuration file that the fuzzer component uses to generate synthetic workloads and inject faults. For Workflow II, a set of traces are provided as input. The simulator drives the nodes and network to operate (in virtual time) by following the actions in a given input trace. Because the actions in the trace (from the model) are not one-to-one with the actions of the simulated implementation, we implement a mapping between the actions as a table that is provided in the configuration file. The checker, after an action is performed, compares the resultant state with the state in the trace and makes sure that they are compatible.  In addition to producing traces for Workflow I, the simulator also measures several performance metrics that can be configured. These measurements are output into a metrics file and used for performance analysis. We note that the deterministic simulator has been implemented so that the node (client code) and its consensus algorithm do not have to be implemented in the same language as the simulator.
\section{Conformance Testing}
\label{sec:testing}
In this section we describe our conformance testing workflows and present an algorithm along with variations that demonstrate how the workflows can be used together. 

\subsection{Overview}
We begin by describing a basic algorithm that combines Workflow I and II in Algorithm~\ref{alg:conftest}. The top-level function \textit{ConfTest} takes as input a formal model $\mathcal{M}$ (as discussed in Section~\ref{sec:formal}, including the model checker) and the implementation running in a deterministic simulator $\mathcal{S}$ (as described in Section~\ref{sec:sim}). The function executes workflows I and II in an alternating manner (indicated by $\vee_{alt}$). If a workflow is successful, the set of successful traces $\aT$ is returned. The returned set of successful traces is used to constrain both $\mathcal{M}$ and $\mathcal{S}$ to avoid regenerating these successful traces. Thus, when $\aT$ is empty, this indicates that all abstract traces have been explored and full coverage has been reached assuming finite traces. 

\begin{algorithm}
 \SetAlgoLined
 \LinesNumbered
 \SetKwInOut{Input}{Input}
 \Input{Formal Model $\mathcal{M}$}
 \Input{Deterministic Simulator $\mathcal{S}$}
 \SetKwProg{Function}{function}{}{end}
 \SetKwRepeat{Do}{do}{while}
 \Function{ConfTest($\mathcal{M}$, $\mathcal{S}$)}{
     \Do{$\aT \neq \emptyset$}{  
     $\aT = \textit{WorkflowI}(\mathcal{M}, \mathcal{S}) \vee_{alt} \textit{WorkflowII}(\mathcal{M}, \mathcal{S})$;
     }
 }

 \Function{WorkflowI($\mathcal{M}$, $\mathcal{S}$)}{
   $\aT = \alpha(\mathcal{S}.fuzz())$\;\label{line:1abstract}
   $\mathcal{S} = \mathcal{S}.constrain(\gamma(\aT))$\;\label{line:1constrained1}
   \uIf{$\mathcal{M}.check(\aT)$}{\label{line:1check}
        $\mathcal{M} = \mathcal{M}.constrain(\aT)$\;\label{line:1constrained}
       \Return $\aT$\;\label{line:1ret}
   }
   \uElse{
     $exit(\mathcal{M}.cex)$\;\label{line:1exit}
   }
 }

 \Function{WorkflowII($\mathcal{M}$, $\mathcal{S}$)}{
   $\aT = \mathcal{M}.fuzz()$\;\label{line:2fuzz}
   $\mathcal{M} = \mathcal{M}.constrain(\aT)$\;\label{line:2constrain1}
   \uIf{$\mathcal{S}.check(\gamma(\aT))$}{ \label{line:2check}
        $\mathcal{S} = \mathcal{S}.constrain(\gamma(\aT))$\; \label{line:2constrain}
       \Return $\aT$\;\label{line:2ret}
   }
   \uElse{
     $exit(\mathcal{S}.cex)$\; \label{line:2exit}
   }
 }
\caption{Model Guided Conformance Testing\label{alg:conftest}}
\end{algorithm}

\subsection{Workflow I}
The function \textit{WorkflowI} takes mutable references to the formal model $\mathcal{M}$ and the deterministic simulator $\mathcal{S}$. In line~\ref{line:1abstract}, $\mathcal{S}$ performs fuzzing and the resulting traces are abstracted to model traces. The simulator is constrained in line~\ref{line:1constrained1} by the traces and the traces are then checked against the model $\mathcal{M}$ in line~\ref{line:1check}. If all the traces pass, we constrain the model in line~\ref{line:1constrained} and return 
$\aT$ in line~\ref{line:1ret}. Otherwise, we exit with the counter-example $cex$ in line~\ref{line:1exit}.

The function $\mathcal{S}.fuzz()$ is implemented as the deterministic simulator fuzzer component. The simulator fuzzes a set of inputs with ranges defined in the 
configuration file. For instance, in our implementation we allow users to configure a subset of the 
inputs in Table~\ref{tbl:params}. The input \texttt{vertex\_production\_rate} specifies how many vertices are being produced per second. Typically, the range is between $0$ and $100$. 
\texttt{num\_nodes} specifies the total number of nodes in the system. The number of faulty nodes is set by \texttt{number\_faulty}, \texttt{failure\_chance} represents the probability that any node crashes. \texttt{failure\_chance} can occur on any Byzantine/faulty node. Inputs \texttt{message\_send\_delay}  and \texttt{message\_receive\_delay} control the network delay that can occur between nodes. \texttt{iteration\_duration} is how long a single iteration of the main loop of the algorithm takes. We configure the fuzzer by defining how many random values to pick from each input (with respect to the ranges). We call this the \emph{fuzzing configuration}. Each configuration is thus associated with a set of traces that are generated e.g., for 3 input variables where 3 random values are selected for each would result in 27 traces. The reason is that, in our implementation, all of the combinations of the values are explored.

\begin{table}[]
\caption{Parameters to Fuzz in Simulator\label{tbl:params}}
\begin{tabular}{l|l|l}
Parameter & Unit & Description \\\hline
vertex\_production\_rate  & int  & vertices produced per second     \\
failure\_chance  &  float    &  Pr[faulty process stops]    \\
number\_faulty &  int    &  number of faulty nodes    \\
num\_nodes &  int    &  node count   \\
message\_send\_delay &  time    & delay for sending      \\
message\_receive\_delay & time &   delay for receiving   \\ 
iteration\_duration & time & time taken for an iteration
\end{tabular}
\end{table}

For each instantiated input combination produced by the fuzzer, a trace is produced. To avoid duplicates, we store the hash of the generated trace for a single fuzzing execution. A mapping file  maps implementation variables to model variables and implementation actions to model actions. This explicit mapping acts as our $\alpha$ relation. The trace is converted to TLA+ and included in the model checker. To check compliance i.e., $\mathcal{M}.check()$ we input the converted trace into an auxiliary TLA+ specification, which serves to execute only the transitions in the input trace and determine if the post-condition state holds for the given initial state and transitions. The trace can be checked as a single model checking problem, or it can be cut up into several sub-traces. We don't observe noticeable performance improvements in either case. 

We constrain the model with all safe traces i.e., $\mathcal{M}.constrain(\aT)$, via rejection sampling \cite{rejectionsample}. This means the trace generator itself operates unconstrained; however, for each trace produced we check if it has already been generated (we store the trace hashes). If the trace has been seen before, the model checker is rerun. This repeats until we get a new trace. This process is efficient because the number of possible traces is very large and the probability of the generator repeating itself is very low in the first place. $\mathcal{S}.constrain(\gamma(\aT))$ is implemented by storing trace hashes permanently for the entire testing session. 


\subsection{Workflow II}
The function \textit{WorkflowII} takes mutable references to the formal model $\mathcal{M}$ and the deterministic simulator $\mathcal{S}$ (as does \textit{WorkflowI}). It obtains traces by fuzzing the model using a model checker in line~\ref{line:2fuzz}. It constrains the model in line~\ref{line:2constrain1} and checks the traces are executable in the simulator in line~\ref{line:2check}. If no counter-example 
is found, the simulator constrains the fuzzer (line~\ref{line:2constrain}) to not produce the safe traces again and returns a set of traces in~\ref{line:2ret}, otherwise it exits with the counter-example $cex$ in line~\ref{line:2exit}. 

We perform  $\mathcal{S}.fuzz()$ by using the TLC model checker's \newline
\texttt{-simulate} command. We can specify a depth $d$ and number $n$ of traces 
for which the model checker performs $n$ random executions of depth $d$.$\mathcal{S}.check(\gamma(\aT))$ is implemented by each trace being ingested by the simulator and the driver stepping through the trace according to the mapping i.e., concreteization function $\gamma$, and checking if each states abstraction is equivalent for the corresponding action(s).

\subsection{Variations}
\label{ssec:var}
Algorithm~\ref{alg:conftest} describes a basic method in which 
 workflows I and II can be combined.  Below we discuss some other variations. 

Firstly, workflows I and II can be performed independently
to find all Type-I and Type-II (mixed with prop) violations. Algorithm~\ref{alg:conftest} showcases that the two problems are complementary (i.e., can share information, can find different violations etc.).  

Secondly, Algorithm~\ref{alg:conftest} terminates when a counter-example trace is found. The algorithm can trivially be modified to continue and find all counter-examples. Furthermore,  Algorithm~\ref{alg:conftest}'s  constraint mechanism is rather simple and only handles constraints based on the trace equality. However, a more elaborate mechanism could perform a subsumption check. This would reduce redundancy but incur a significant computation cost. We also note that while \textit{ConfTest}'s main loop theoretically terminates when $\aT$ becomes empty (for finite models with finite traces), in practice we typically bound the number of loop iterations and do not place bounds on the model size. 


Next, Algorithm~\ref{alg:conftest} can easily be adapted to perform the feedback loop between two implementations or two models. This is interesting when equality between protocols in two implementations or two models needs to be established. While equality of models can in practice be established via proof, equality of implementations is a harder endeavor and such a variation has utility. The simulation thus far assumed a single protocol is simulated on multiple nodes. However, the simulator can support the simulation of several implementations side by side.








\section{Experimental Evaluation}
\label{sec:evaluation}
In this section we evaluate our framework in several areas, including 
precision, performance and manual effort. 

\subsection{Experimental Setup}
MacOS 14.1 on an Apple M3 with 18 GB memory. We use 
TLA+ toolbox version 1.8.0 with TLAPS version 1.5. and golang version go1.23.2 darwin/arm64.

We perform our analysis on a DAG-based consensus protocol as described in Section~\ref{sec:usecase}. The consensus protocol implementation is written in golang and consists of 2411 lines of code. The deterministic simulator consists of 1000 lines of code for the core driver and another 1000 lines of code for the network/DES abstractions. The specification is written in TLA+ and consists of 675 lines of code and 1282 lines of proof code. 

\subsection{Violation Discovery Evaluation}
In the following experiments, we evaluate our framework on a number of historic conformance violations in the code base (both implementation and specification). We re-injected each violation into our correct implementation and specification 
and evaluated the fuzzing configuration required for the respective workflow to discover the violation. The configuration for 
Workflow I is a tuple identifying how many variables were fuzzed and how many random values were drawn for each 
variable. We generally use 3 variables, namely, number of nodes, iteration duration (in milliseconds) and chance of 
node failures. Thus, 3/1 can be interpreted as all three variables having 1 random value drawn each, resulting in 1 
trace; 3/2 can be interpreted as all three variables having 2 random value drawn each. The minimum is 3/1 the maximum is 3/3 (one could go further to 3/4 etc. but we stopped here in our testing). 
Depending on the algorithm (as described in Section~\ref{sec:testing}) each configuration could be interpreted as several executions of the workload in a loop or the workload was executed for several traces (see Section~\ref{ssec:var}). For Workflow II, the configuration is  the number of traces generated and their depth, e.g., 2/5 means 2 traces of depth 5. The minimum configuration we try is 10/1000 and the maximum is 500/1000. For all violations,  we report the minimum configuration that found the error. 

In Table~\ref{tbl:precision}, we show 10 violations (Violation) that we re-injected. For each we define the type, where the fix occurred (Fix), the concrete values that produced the trace that detected the violation (Value), the minimum configuration value to detect the violation (Min. Config.), the length of said trace (Len.) and the time to generate (T-Gen.) and verify (T-Ver.) all traces (including additional correct traces). The value is a triple of number of nodes, iteration duration in milliseconds and probability that nodes can be faulty. 

Overall, detection for most (60\%) of violations took less than a minute, with others (40\%) requiring between 1-10 minutes.  Factors in detection time 
are the number of traces required to detect the violation, the length of the trace and the location in the trace of the violation. For our use case, we found that Workflow I is more expensive but provides more utility (more exclusively Type-I violations, more control over trace generation etc.). The performance for our benchmarks and protocol was sufficient, however finding deeper more inhibitive violations may require more time if more iterations are needed or more traces need to be generated at once. Below we describe each violation in Table~\ref{tbl:precision} and comment on the results for each violation. 


\begin{table*}[]
\caption{Detection of historic conformance violations in our consensus protocol \label{tbl:precision}}
\begin{tabular}{l|l|l|l|l|l|l|l}
Violation & Type & Fix     & Value                           & Min. Config. & Len.    & T-Gen. & T-Ver.   \\ \hline
1 & Type-I   &  Spec.  & $\langle 7, 25.31, 0.54 \rangle$     & 3/1   &  1214   &   10.34s  & 7s           \\
2 & Type-I  &  Spec.   &   $\langle 17, 26.4, 0.07 \rangle$   & 3/1   &  2718   &   18.98s & 6s            \\
3 & Type-I  &  Impl.   &   $\langle 11, 17.43, 0.23 \rangle$  & 3/2   &  1901   &   124.5s &  10m6s          \\
4  & Type-I  & Spec.   &   $\langle 8, 25.42, 0.37 \rangle$   & 3/1   &  1546  &   7.2s   &  1m45s        \\
5 & Type-II  & Impl.   &  -         & 10/1K                            &   1000  &    3s     &   48.5s        \\
6 & Prop.   &  Impl.   &   $\langle 13,21.75,0.14 \rangle$   & 3/2    &  2662   &    95.82s  & 4m32s  \\ 
7 & Type-I &   Impl.   &  $\langle 7,11.24,0.89 \rangle$     & 3/1    &  1271   &   8.65s    & 1m4s        \\
8 & Type-I  &  Impl.   &  $\langle 20,17.11,0.35  \rangle$  & 3/1     &  1162   &    26.87s   &  7s             \\
9 & Type-I  &  Impl.   &  $\langle 14,18.62,0.27\rangle$    & 3/1     & 2052   &    12.67s   &  6s             \\
10 & Type-II  &  Spec. &  -                                 & 10/1K   &  1000    &    3s   &  17.2s             \\
\end{tabular}
\end{table*}

\paragraph{\bf Type-I Violations} 
We summarize the results of the Type-I violations found below.
\begin{description}[style=unboxed, leftmargin=0cm]
    \item[Violation 1:] This violation is a round numbering discrepancy where 0 indexing was used instead of 1 indexing. The fix could have been done in either the model specification or the implementation, but we opted to fix it in the specification. The violation was found with single trace from the values $\langle 7, 25.31, 0.54 \rangle$. The detection took 17 seconds.
    \item[Violation 2:] This violation occurred because the genesis (initial) vertex was not included in wave 1 in the specification. While this was a specification error, it did not affect 
    safety or the proof. We fixed the specification to align with the implementation.  The trace that discovered the violation had the values $\langle 17, 26.4, 0.07 \rangle$. It was discovered in 25 seconds.  
    \item[Violation 3:] This violation occurred because in the implementation one vertex can (rarely) be received two times by the same node. For this violation we needed to vary the chosen variables into 2 values each (i.e. use 3/2 configuration) as the first trace generated via 3/1 did not detect the violation. Instead, trace $\langle 11, 17.43, 0.23 \rangle$ detected the violation. The detection took 12 minutes and 10 seconds. 
    \item[Violation 4:] This violation was a leader logic error where the specification assumes sentinel values for values that are yet to be determined, while the simulator does not. We changed the specification to conform to the implementation. The trace that found the violation was generated with $\langle 8, 25.42, 0.37 \rangle$ and the generation and node took 1 minute and 52 seconds. 
    \item[Violation 7:] This violation highlighted an underflow bug in the implementation that changed the round and election logic. It was detected with values $\langle 7,11.24,0.89 \rangle$ in 1 minute and 13 seconds. 
    \item[Violation 8:] This violation highlighted that in the implementation nodes could receive vertices from next rounds before creating the self ancestor, which is forbidden in specification. We detected this violation with $\langle 20,17.11,0.35  \rangle$ in 34 seconds. 
    \item[Violation 9:] In this violation nodes did not increase their round numbers in time which made them behave as if they were not in their proper round. We detected this with values $\langle 14,18.62,0.27\rangle$ in 19 seconds. 
\end{description}

\paragraph{\bf Type-II Violations} We summarize the results of the Type-II violations found below.

\begin{description}[style=unboxed, leftmargin=0cm]
\item[\textbf{Violation 5:}] The simulator did not support a change of number of nodes after a constant number of rounds. We detected this violation in 52 seconds and changed the implementation to contain this logic. 
\item[\textbf{Violation 10:}] In this violation the implementation did not support equivocation as it assumed it cannot occur. However the specification despite assuming reliable broadcast modeled equivocation. We detected this violation in 20 seconds and changed the specification. 
\end{description}

\paragraph{\bf Property Violations} We summarize the results of the property violations found below.

\begin{description}[style=unboxed, leftmargin=0cm]
\item[\textbf{Violation 6:}] Method of ordering was technically not ordered as assumed and it occasionally violated consistency. As with Violation 3, we needed a 3/2 configuration, wherein  $\langle 13,21.75,0.14 \rangle$ discovered the violation. The detection took 6 minutes. 
\end{description}

We note that several of the Type-I violations could be classed as Type-II violations from the view of model traces (1, 2, 4, 7), however several can only be classed Type-I (3, 8, 9) or Type-II (5, 10). Since we started with Workflow I there was a bias to find Type-I violations first.

\subsection{Detailed Performance Evaluation}
\begin{figure}[!h]
        \centering
    \begin{subfigure}{.487\linewidth}
            \centering
  \begin{tikzpicture}
        \begin{axis}[                    
            width=1.15\textwidth,
            xtick={1,2, 3, 4, 5, 6, 7},
            scaled ticks=false,
            xticklabels={1/1, 1/2, 1/3, 2/2, 2/3, 3/2, 3/3},
              ylabel near ticks,
        legend pos=south east,
            ]
            \addplot coordinates {
               (1, 207)
                (2, 871)
                (3, 1665)
                (4, 1950)
                (5, 5219)
                (6, 3716)
                (7, 14434)
                };
        \end{axis}
    \end{tikzpicture}
    \caption{Configurations vs Workflow I time (sec)\label{fig:wf1}}
\end{subfigure}\hfill
\begin{subfigure}{.487\linewidth}
            \centering
  \begin{tikzpicture}
        \begin{axis}[                    
            width=1.15\textwidth,
            scaled ticks=true,
            xtick={1, 2, 3},
            xticklabels={10/1K, 100/1K, 500/1K},
              ylabel near ticks,
        legend pos=south east,
            ]
            \addplot coordinates {
                (1, 123)
                (2, 1254)
                (3, 6270)
            };
        \end{axis}
    \end{tikzpicture}
    \caption{Configurations vs Workflow II time (sec)\label{fig:wf2}}
\end{subfigure}
\caption{Run-time of Workflows I and II for representative configurations \label{fig:wf12}}
\end{figure}
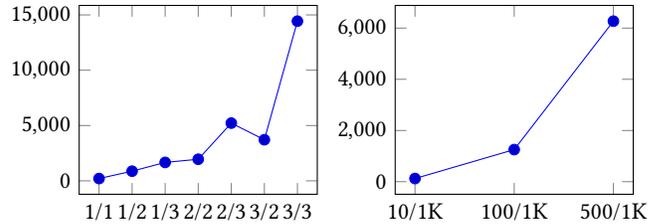

We perform a further performance breakdown in Figure~\ref{fig:wf12} of each workflow for various representative configurations. Since the larger configurations generate more safe traces, the run-time which combines generations and verification times increases. This evaluation provides an insight into the trade-offs between 
the different setups of the workflows. Depending on the use case, the times may or may not be inhibitive. For instance, if testing is performed offline in the background, larger run-times may be acceptable. On the other hand, for more interactive use cases (e.g., CI/CD pipeline), single trace generation per workflow (like in our use case) may be preferred. Our evaluation generates only safe traces to represent the worst case. However, we note that when violations are found, the trace checking terminates earlier and improves run-time. 

For Workflow I (Figure~\ref{fig:wf1}), the run-times range from approx. 3 minutes to 4 hours. For Workflow II (Figure~\ref{fig:wf2}),  the run-times 
range from approx. 2 minutes to approx. 2 hours. The run-time 
results further motivate the need for fuzzing since 
verification via model checking for our protocol specification does not scale. 

\subsection{Manual Effort}
The effort to design, understand, specify and prove the protocol required approximately 2 person-months, distributed between 3 people. The people were familiar with formal methods/logic but had to learn TLA+. This is a similar time period required to produce a typical design document, and the specification acted as an artifact that is referenced when ambiguity is encountered. The time for the proof to be machine-checked took 123 seconds (approx. 2 minutes).




\section{Discussion}
\label{sec:dis}
In this section we discuss our framework including its limitations and lessons learnt 
applying it to our industrial use case. 

\paragraph{\bf Limitations and Threats to Validity} In 
our experiment, we evaluate our framework on our use case, namely, a blockchain consensus protocol. Therefore, 
our experimental results may not reflected in other systems. However, 
considering our specification could not be model checked in a practical amount of time and space, we believe our protocol represents a particularly difficult case. We therefore argue that our framework will have even better performance on simpler protocols and unlock conformance testing for other protocols that are hard to verify with model checkers.  We leave the application of our framework on more diverse systems for future work. 

Our implemented framework uses TLA+ which requires a background in formal logic. Since we have a formal methods background, this may introduce implicit bias towards the individual developers’ expertise. In future, more user studies with other developers (e.g., average distributed system’s developer) could further validate or challenge our approach.

\paragraph{\bf Lessons Learnt}
Our project started with the observation that our specification in TLA+ was 
impossible to model check for realistic instances and existing techniques that could be used to establish partial conformance e.g.,\cite{WangDGWW023},
did not scale for realistic instances e.g., >3 nodes. Thus, our first lesson learnt 
was that model checking is not a one-size-fits-all approach and  protocols such as ours can benefit from combining fuzzing, model checking with interactive proof. 

Another lesson learnt 
was the processes of establishing 
a connection between model and implementation is not trivial. We found it is much easier to provide a mapping with an initial specification that is fairly abstract. Later, the specification can be refined to include more detail. However, developing a specification from an existing implementation is much harder. 

Lastly, our testing approach discovered Type-I and Type-II violations, which resulted in us refining \emph{both} the specification and implementation.  While the violations were mostly fixed in the implementation, some resulted in changing the specification, which highlights that the specification is not static and can evolve particularly during initial implementations.

\section{Related Work}
\label{sec:relatedwork}
In this section we compare our technique to relevant related work. The scope of testing 
general distributed systems is very extensive, therefore we focus on techniques that share 
aspects of our approach. 

\paragraph{\bf Model-based testing for distributed systems.}
The work in~\cite{conftest1} proposes a conceptual unidirectional testing framework for protocol conformance. It doesn't outline any 
particular specification language or tooling, rather it outlines how tests can be generated from a mathematical description of a system. The technique resembles a unidirectional version (only Workflow II) of our framework.  The work in~\cite{WangDGWW023} present \textsc{Mocket} which extracts \emph{all} valid traces of a \emph{finite} model during model checking and executes the behavior in a reference system that is not necessarily deterministic. Our attempt at employing \textsc{Mocket} on an instance with 4 nodes and 4 rounds on our protocol specification did not scale (ran out of memory -- 18 Gig). Even employing it on a partial specification (only DAG construction) with a small instantiation (3 nodes and 2 rounds) resulted in a several Gigabyte dot file that took many hours to extract. For specifications where \textsc{Mocket} scales, the technique can replace the need for fuzzing in Workflow II, and hence is very complementary to our technique. Most notably since this technique is only comparable to our 
Workflow II, as it can miss added unwanted behavior in the implementation (or missing in specification) that is detected by Workflow I. 

Alternatively, the technique in~\cite{CirsteaKLM24} is similar to our Workflow I. However,  it does not perform fuzzing, but rather log executions during execution. Thus this technique 
resemble execution-time monitors and misses violations that can only be found using Workflow II.  Similarly, the technique in~\cite{FooCC23} considers modeling a TLA+ specification and compiling it down to a monitor in the implementation language for each local node. For this they perform a projection on the TLA+ specification to extract local node behavior. This however limits the checking to conformance testing on local node behavior and cannot check for network-centric properties which are vital for blockchains. Moreover, since they deploy (monitor) in 
a real-world system they cannot test for a wide range of behaviors in a short amount of time. Since these techniques are only comparable to our 
Workflow I, they can miss added missing behavior in the implementation (or additional in specification) that is detected by Workflow II. 

\paragraph{\bf Deterministic simulation testing.} Deterministic simulation testing was introduced by  \textsc{FoudnationDB}~\cite{fdb} which provided a deterministic simulator alongside the \textsc{FoudnationDB} database. The obvious difference is that the simulator in~\cite{fdb} is designed specifically for the \textsc{FoudnationDB} database. Our simulator on the other hand is made for Blockchains i.e., it's composed of different components and ingests different inputs. On a conceptual level, unlike our approach, \cite{fdb} requires manual placement of assertions to check only local properties. Moreover, these "test oracles" are ad hoc and are limited to simple properties. Our approach allows for both local and global properties and does not require any manual placement into source code. Checking is performed via a model checker and a model. 



\section{Conclusion}
\label{sec:conclusion}
We have presented a novel conformance testing technique and implemented it as part of the FMDSE framework. The technique consists of two complementary workflows that can be configured algorithmically in several ways. We demonstrate the utility of our framework on an industrial strength consensus protocol and show that a number of real world conformance violations can be detected. Most notably, we should note that for a full coverage of violation types, both workflows are required. While in this paper we have focused on our use case, namely, the conformance of a consensus protocol, we believe our framework has utility for conformance-based testing of distributed systems in general.

\bibliographystyle{abbrv}
\bibliography{main} 

\end{document}